\documentclass[12pt]{article}
\usepackage{subfigure}
\usepackage{url}
\usepackage{graphicx}
\usepackage{algorithmic}
\usepackage{algorithm}
\usepackage{acronym}
\usepackage{xspace}
\usepackage{xcolor}
\usepackage{amsmath}
\usepackage{amssymb}
\usepackage{setspace}
\acrodef{BWI}[BWI]{Bell-Wigner Inequality}
\acrodef{IAR}[IAR]{Information Access and Retrieval}
\acrodef{IR}[IR]{Information Retrieval}
\acrodef{LSA}[LSA]{Latent Semantic Analysis} 
\acrodef{NPL}[NPL]{Neyman-Pearson Lemma}
\acrodef{PRP}[PRP]{Probability Ranking Principle}
\acrodef{QIRB}[QIRB]{Quantum Information Retrieval Basis}
\acrodef{QM}[QM]{Quantum Mechanics}
\acrodef{SVD}[SVD]{Singular Value Decomposition}
\acrodef{WWW}[WWW]{World Wide Web}
\acrodef{QM}[QM]{Quantum Mechanics}
\acrodef{QPS}[QPS]{Quantum Probability Space}
\acrodef{ML}[ML]{Machine Learning}
\acrodef{QLRA}[QLRA]{Quantum-Like Representation Algorithm}
\acrodef{IR}[IR]{Information Retrieval}
\acrodef{NPL}[NPL]{Neyman-Pearson's Lemma}
\acrodef{BWI}[BWI]{Bell-Wigner Inequality}

\newcommand{\ket}[1]{{#1}}
\newcommand{\bra}[1]{\ket{#1}^{\prime}}

\newcommand{\tr}{\mbox{tr}}

\title{An Algorithm to Calculate a\\\acl{QPS}}
\author{
  Massimo Melucci\\
  {University of Padua, Italy}\\
  \url{massimo.melucci@unipd.it}
}
\date{}

\begin{document}
\maketitle

\section{Introduction}
\label{sec:introduction}

Many scientific disciplines need probability spaces to fit observed
data, predict future data or explain relationships.  To this end, a
\emph{probability space} should be utilized to represent events by
sets and leverage the closure under conjunction, disjunction and
negation \cite{Kolmogorov56} to examine more complex events according
to the tie with the Boolean logic \cite{Boole1854}.

Context\footnote{Context origins from Latin \textit{contextus}, from
  \textit{con}-'together' + \textit{texere} 'to weave'.}  is viewed as
the complex of experimental conditions in which uncertain and related
events are observed in terms of data.  Kolmogorov implicitly assumes a
context when he refers to ``a complex of conditions which allows of any
number of repetitions'' \cite[page 3]{Kolmogorov56}.  In this paper, we
explicitly state that a probability space is the mathematical
representation of a context because the space provides a representation
of contextual events, their relationships and measures of the
uncertainty affecting the observed and the predicted data.
The uniqueness of the context from which data are observed is crucial
when an event is predicted conditionally to the observation of other
events.  As a context coincides with one probability space, the
prediction of the event conditionally to the other events would be
possible if the observed data could define a single probability space
and then only one context.


In this paper we address the problem of using one probability space
for estimating parameters and predicting future data when the observed
data come from multiple contexts and thus from distinct spaces.  We
explain that a set-based probabilistic space might be suboptimal in
the case of multiple contexts.  To overcome suboptimality and
reconcile multiple contexts in one space, the paper introduces the
\ac{QPS} whose foundations were described in \cite{Feynman&65}.  We
also present an algorithm to calculate the \ac{QPS} for data observed
from multiple contexts and provide a web application that implements
the
algorithm\footnote{\url{http://isotta.dei.unipd.it/cgi-bin/qps/qps-w-form.py}}.


\section{Probability Spaces}
\label{sec:probability-spaces}

In probability, a random experiment is an experimental context where
the data are observed in conditions of uncertainty. As the events need
a numerical representation, a random variable is a function mapping an
event observed in a random experiment to a point of an interval of the
$n$-dimensional data space; for example, a binary variable maps coin
toss to $\{0,1\}$.  Random variables provide a succinct and sufficient
description of events because the events mapped to a certain data are
in the same subspace; for example, the movies a person does (or does
not) like are in the same subspace labeled by $A$ (or respectively
$\bar{A}$) including all the movies that the user likes (or does not
like); a random variable maps $A$ (or $\bar{A}$) to $1$ (or $0$).

A probability space is the mathematical representation of a random
experiment.  Formed by a set of events and by a probability function,
or density operator, a probability space assigns a probability to each
subspace with three fundamental properties.  The empty subspace is
mapped to $0$, the whole space is mapped to $1$, and for any pairwise
disjoint subspaces, the probability of the disjunction is the sum of
the probabilities assigned to each subspace. (Two subspaces are
disjoint when one is included in the orthogonal subspace of the
other.) 

Suppose $A_1,A_2,A_3$ be three subsets corresponding to $n=3$ binary
variables that assign an event to a subset if and only if the value
observed for the corresponding variable is one.  After measuring the
three variables for a certain sample, the marginal probabilities like
$P(A_2A_3)$ are available to calculate $P(A_1A_2A_3) = P(A_1|A_2A_3)
\, P(A_2A_3)$; however, the number of subsets is $2^n$, thus making
the calculation of probability for any number of variables infeasible
for non-small $n$.  \emph{Conditional independence} can be a helpful
workaround for overcoming the problem of the exponential number of
subsets.  Two variables $A_2$ and $A_3$ are conditionally independent
on $A_1$ when
\begin{equation}
  \label{eq:cond-indip}
  P(A_2A_3|A_1)=P(A_2|A_1)\,P(A_3|A_1)
\end{equation}

Conditional independence requires marginal probability values. Howver,
a set-based probability space implies some statistical inequalities
between the marginal probability values.  Therefore, the violation of
an inequality implies the inexistence of a single set-based
probability space.  For example, suppose that the following marginal
probabilities are provided by some distinct contexts:
\begin{equation}  
  P(A_1A_2)=\frac{9}{20} \qquad P(A_1A_3)=\frac{9}{20} \qquad P(A_2A_3)=\frac{1}{10}
  \label{eq:bwi-false}
\end{equation}
\begin{equation}
  P(\bar{A}_1)=\frac{1}{2} \qquad P(\bar{A}_2)=\frac{1}{2} \qquad P(\bar{A}_3)=\frac{1}{2}
  \nonumber
\end{equation}

A probability space cannot exist for $A_1,A_2,A_3$ because no
$P(A_1A_2A_3)$ can be calculated; otherwise, we would have to accept
sets of negative volume.  The situation is similar to Euclid's theorem
according to which the inner angles of any triangle shall sum to $\pi$
only if placed on a plane.  If placed on a type of surface other than
a plane, the angles of the triangles of stars might not sum to $\pi$.
If the observed angles violates Euclid's theorem, planarity does not
hold \cite{Accardi84}.

A test of existence of a single probability space for any $n$ is
provided in \cite{Pitowsky88}.  Let $A_1,A_2,A_3$ be three binary
variables.  Suppose we are provided with $P(A_1)$, $P(A_2)$, $P(A_3)$,
$P(A_1A_2)$, $P(A_1A_3)$, $P(A_2A_3)$ from distinct experimental
contexts.  Let
\begin{footnotesize}
  \begin{equation}
    \ell = \max\{0, P(A_1A_2)+P(A_1A_3)-P(A_1),  P(A_1A_2)+P(A_2A_3)-P(A_2), P(A_1A_3)+P(A_2A_3)-P(A_3)\}
    \nonumber
  \end{equation}
\end{footnotesize}
and
\begin{footnotesize}
  \begin{equation}
    \upsilon = \min\{P(A_1A_2),P(A_1A_3),P(A_2A_3),1-(P(A_1)+P(A_2)+P(A_3)-P(A_1A_2)-P(A_1A_3)-P(A_2A_3))\} 
    \nonumber
  \end{equation}
\end{footnotesize}
The following inequality holds if $A_1,A_2,A_3$ refer to the same space:
\begin{equation}
  \ell \leq P(A_1A_2A_3) \leq \upsilon
  \label{eq:inequality}
\end{equation}
In other words, one context would be possible only if
\eqref{eq:inequality} held.  Although there are other inequalities to
consider \cite{Accardi&82}, there is a general result that holds for
every $n$ \cite{Pitowsky88}.

In other words, if \eqref{eq:inequality} does not hold, then
$A_1A_2A_3$ cannot be defined, and distributivity, that is
\begin{equation}
  A_1(A_2 \lor A_3)=(A_1A_2) \lor (A_1A_3)=(A_1A_2\bar{A}_3) \lor ({A_1A_2A_3}) \lor(A_1\bar{A}_2A_3)
  \nonumber
\end{equation}
cannot hold. Since distributivity is a feature of sets, eventually one
set-based probability space does not exist for all the variables
$A_1,A_2,A_3$ when \eqref{eq:inequality} does not hold.

One approach to calculating $P(A_1A_2A_3)$ when $A_1A_2A_3$ cannot
exist is to utilize conditional independence \eqref{eq:cond-indip}.
In this way, $P(A_1A_2A_3)$ can be approximated even though
$A_1A_2A_3$ cannot exist. However, some efficiency is lost when using
conditional independence as explained in Section \ref{sec:prob-rank}.

\section{Probabilistic Ranking}
\label{sec:prob-rank}

Let $A_1,A_2,A_3$ be three binary variables corresponding to three
subspaces.  A decider has to split the set of observed values for
$A_1,A_2,A_3$ in an acceptance region (${\cal A}$) and its complement;
the acceptance region is the subset of triples of binary values such
that the decider takes one out of two options; for example, a
classifier decides for each class whether a triple of values observed
for an event is in the acceptance region and therefore if the event
should be put in a class.  Let
\begin{equation}
  \nonumber
  P_i({\cal A})=\sum_{A_1,A_2,A_3 \in {\cal A}} P_i(A_1A_2A_3) 
\end{equation}
be the likelihood of class $i$ given the observed data.  The system's
decision is optimal when the set of values is split in such a way as
to maximize $P_1({\cal A})$ while keeping the likelihood of the
complement class small.  

The \ac{NPL} states that optimal decision can be obtained when
$P_1({\cal A}) - cP_0({\cal A})$ is maximum provided that $P_0({\cal
  A}) \leq \alpha$ \cite{Neyman&33}.  While varying $c$, the decider
produces a ranking; at the top of the ranking, the decider puts the
first items, while the least preferred items are ranked at the bottom
of the ranking.

However, optimal decision require the existence of $A_1A_2A_3$ which
cannot be taken for granted if there are distinct experimental
contexts, although it might not be calculated nor approximated by
conditional independence.

Suppose $A_1,A_2,A_3$ are measured in three distinct contexts
corresponding to the pairs $(A_1,A_2)$, $(A_1,A_3)$ and $(A_2,A_3)$,
of which marginal probabilities may violate \eqref{eq:inequality}.
Suppose the system ranks items by assuming conditional independence
\eqref{eq:cond-indip}.  If following marginal probabilities are
estimated
\begin{equation}
  P(A_1A_2)=\frac{1}{4} \qquad P(A_1A_3)=\frac{1}{4} \qquad P(A_2A_3)=\frac{1}{4}
  \nonumber
\end{equation}
\begin{equation}
  P(\bar{A}_1)=\frac{1}{2} \qquad P(\bar{A}_2)=\frac{1}{2} \qquad P(\bar{A}_3)=\frac{1}{2}
  \nonumber
\end{equation}
the following triples, $\bar{A}_1A_2A_3$, $A_2\bar{A}_1\bar{A}_3$,
$\bar{A}_1\bar{A}_2A_3$, $\bar{A}_1 \bar{A}_2 \bar{A}_3$,
$A_1A_2\bar{A}_3$, $A_1\bar{A}_2A_3$, $A_1\bar{A}_2\bar{A}_3$ will be
equally ranked.  According to \eqref{eq:inequality}, $P(A_1A_2A_3)$
ranges between $\ell=0$ and $\upsilon = \frac{1}{4}$; therefore,
$A_1A_2A_3$ may be ranked before, after or coincidentally with any
other triple depending on the probability space and provided the same
marginal probabilities.  As there may be an infinity of $P(A_1A_2A_3)$
satisfying \eqref{eq:inequality} provided the aforementioned marginal
probabilities, the approximation of $P(A_1A_2A_3)$ by
$P(A_2|A_1)P(A_3|A_1)$ is only one out of the infinity of admittable
values whereas the true and unknown value of $P(A_1A_2A_3)$ might be
in any of the points of the interval $[\ell,\upsilon]$.  As
$P(A_2|A_1)P(A_3|A_1)$ is only one out of the infinity of admittable
values, the ranking resulting from the assumption of conditional
independence may place $A_1A_2A_3$ on the top or on the bottom, thus
causing suboptimal ranking.  To overcome suboptimality, $P(A_1A_2A_3)$
should be calculated and not only approximated through conditional
independence.  However, when the number of variables is not small, the
number of probability calculation becomes exponentially large.

In the next two sections we show that the \ac{QPS} calculates
$P(A_1A_2A_3)$, avoids suboptimality and optimally ranks events even
though \eqref{eq:inequality} is violated.

\section{The \acl{QPS}}
\label{sec:qps}

Sets are not the only framework of a probabilistic space.  Instead of
sets, the \ac{QPS} utilizes vectors and the operators thereof.  In
this section, we first define the \ac{QPS} in terms of subspaces and
probability function; then, we introduce an algorithm to compute the
subspaces and the probability function.

\subsection{Definition of the \ac{QPS}}
\label{sec:definition-qps}

Let ${\cal H}$ be a vector space and $\ket{x}\in{\cal H}$ be a vector,
i.e. the simplest subspace.  Let $\ket{0}$ be the null vector. The
join $\ket{x} \oplus \ket{y}$ is the smallest subspace including both
$\ket{x}$ and $\ket{y}$.
Let $Q$ be a probability function; the \ac{QPS} is given by ${\cal H}$
and $Q$ where $Q(\ket{x}), \ket{x}\in{\cal H}$ is the probability of
$\ket{x}$ and
\begin{enumerate}
\item $Q(\ket{0})=0$
\item $Q({\cal H})=1$
\item $Q(\ket{x} \oplus \ket{y})=Q(\ket{x}) + Q(\ket{y})$ for any pair
  of disjoint subspaces $\ket{x}$ and $\ket{y}$.
\end{enumerate}
We label the output probability space as ``quantum'' because it is
based on the mathematical formalism of \acl{QM}.  For this reason, we
use $Q$ instead of $P$, which is left for set-based probability
spaces.

Essential to the \ac{QPS} is Gleason's theorem \cite{Gleason57}.  If
an event is represented by a vector $\ket{x}\in{\cal H}$, the
probability function $Q(\ket{x})$ is necessarily a density matrix
$\rho$ such that
\begin{equation}
  Q(\ket{x})=\bra{x}\rho\,\ket{x}
  \label{eq:gleason}
\end{equation}
is a bilinear quadratic form.  The hard part of Gleason's theorem is
calculating $\rho$.  In this paper, we explain an algorithm to the aim
of calculating $\rho$ that can reproduce the marginal probabilities
and calculate the probabilities of the events when they cannot be
expressed by the set-based space.

The \ac{QPS} can measure variables outside the framework based on
sets.  In the framework based on vector spaces, $Q$ may still be
calculated even though the marginal probabilities violate
\eqref{eq:inequality}.  As a consequence, a single \ac{QPS} can be
defined although the marginal probabilities come from distinct
contexts.  Of course, the probabilities provided by the \ac{QPS} might
differ from those provided by a set-based space even though both
spaces provide the same marginal probabilities.  Therefore, the
\ac{QPS} may give another ranking of items.  Whether this alternative
ranking is better than the ranking provided by a set-based space is a
matter of experimentation.

\subsection{An Algorithm to Calculate the \ac{QPS}} 
\label{sec:an-algorithm-calculate}

In this section, we explore the search for a unique probability space
to move to a theoretical framework other than Kolmogorov's.  The basic
idea is that optimality of ranking may be recovered if
\eqref{eq:inequality} can be violated.
To this end, the algorithm introduced in this paper takes marginal
probabilities as input and gives one \ac{QPS} as output, although the
variables are measured from distinct experimental contexts and may
violate \eqref{eq:inequality}.

In general, the problem is as follows.  Suppose that there are $n$
binary variables $A_{1},\dots,A_n$ and $m = n(n+1)/2 $ univariate or
bivariate marginal probabilities
\begin{equation}
  \nonumber
  P(\bar{A}_i) \qquad P(A_iA_j) \qquad i=1,\dots,n-1 \qquad j=i+1,\dots,n
\end{equation}
Let $b_n=i_{1}\,i_{2} \cdots i_n \in \{0,1\}^n$ be a $n$-digit binary
string\footnote{When $n=0$, $b_{0}$ is an empty string.}.  We have
that the event
\begin{equation}
  A_{1}=i_{1},\dots,A_n=i_n
  \nonumber
\end{equation}
can be shortly written in terms of canonical vectors of $\{0,1\}^n$ as
follows:
\begin{equation}
  \nonumber
 A_{b_n} = i_1 \cdots i_n
\end{equation}
The algorithm calculates $\rho$ such that
\begin{align*}
  P(\bar{A}_i)	&=\left(\bigoplus_{b\in b_{i-1}0 b_{n-i}} \bra{x}_b \right) \,\rho\, \left(\bigoplus_{b\in b_{i-1}0 b_{n-i}} \ket{x}_b\right) \\ i=1,\dots,n\\
  P(A_iA_j)	&=\left(\bigoplus_{b\in b_{i-1}1b_{j-i}1b_{n-j-1}}
    \bra{x}_b\right) \,\rho\, \left(\bigoplus_{b\in b_{i-1}1b_{j-i}1b_{n-j-1}} \ket{x}_b\right) \\ 
  i=1,\dots,n-1 \\ j=i+1,\dots,n 
\end{align*}
and
\begin{equation}
  \nonumber
  0 \leq \bra{x}_b\,\rho\,\ket{x}_b \leq 1
  \qquad \sum_{b \in \{0,1\}^n} \bra{x}_b\,\rho\,\ket{x}_b=1
\end{equation}
particular, a generic element of the $\Lambda$'s diagonal is defined
as
\begin{equation}
  \nonumber
  \lambda_i=
  \left\{
    \begin{array}{ll}
      P(\bar{A}_i)&i=1,\dots,n
      \\
      P(A_kA_j)&i=(k-1)n+j \quad k=1,\dots,n-1 \quad j=k+1,\dots,n   
    \end{array}
  \right.
\end{equation}
\begin{figure}
  \begin{algorithmic}[1]
    \REQUIRE $n$ binary variables $A_{1},\dots,A_n$ such that either $\bar{A}_i$ or $A_i$.  
    \REQUIRE $P(\bar{A}_i)$ for $i=1,\dots,n$.  
    \REQUIRE $P(A_iA_j)$ for $i=1,\dots,n-1$ and $j=i+1,\dots,n$.  
    \STATE{$m \gets \frac{n(n+1)}{2}$} 
    \STATE{$N \gets 2^n$}
    \STATE\COMMENT{\color{gray}Build the $m \times N$ matrix $K$ as
      follows:\color{black}} 
    \FORALL{$i=1,\dots,n$}{
      \STATE\COMMENT{\color{gray}Fill row $i$ by alternating $2^{{n-i}}$ ones and $2^{{n-i}}$ zeros.\color{black}} }
    \ENDFOR
    \STATE\COMMENT{\color{gray}Fill the remaining $\frac{n(n-1)}{2}$ rows as follows:\color{black}} 
    \FORALL{$\ell=1,\dots,N$}{ \STATE{$k
        \gets n$} \FORALL{$i=0,\dots,n-1$}{ \FORALL{$j=i+1,\dots,n$}{
          \IF{$K[i,\ell]=K[j,\ell]=0$}{ \STATE $K[k,\ell] \gets 1$ } \ELSE{
            \STATE $K[k,\ell] \gets 0$ }
          \ENDIF
          \STATE{$k \gets k+1$} }
        \ENDFOR
      }
      \ENDFOR
    }
    \ENDFOR
    \FORALL{$i=1,\dots,n$}{ \STATE $\lambda_{i} \gets P(\bar{A}_i)$ }
    \ENDFOR
    \FORALL{$i=1,\dots,n-1$}{ \FORALL{$j=i+1,\dots,n$}{ \STATE $k \gets
        \frac{1}{2}\left({n\,\left(n-1\right)}-{\left(n-i\right) \,\left(
              n-i-1\right) }\right)+j$ \STATE $\lambda_{k} \gets
        P(A_iA_j)$ }
      \ENDFOR
    }
    \ENDFOR
    \STATE\COMMENT{\color{gray}Compute \eqref{eq:j}\color{black}}
    \STATE\COMMENT{\color{gray}Compute \eqref{eq:spectral-theorem}\color{black}}
    \STATE\COMMENT{\color{gray}Compute \eqref{eq:r}\color{black}}
    \STATE\COMMENT{\color{gray}Compute \eqref{eq:rho}\color{black}}
  \end{algorithmic}
  \caption{An algorithm for computing the density matrix given the
    marginal probabilities of $n$ binary variables.}
  \label{algorithm}
\end{figure}

\noindent The general algorithm can be described in Figure
\ref{algorithm}.  First, a $n \times N$ binary matrix $K$ is generated
(lines 1--20).  The algorithm proceeds with lines 21--29 where the
input marginal probabilities are arranged in a diagonal matrix
$\Lambda$. Finally, the matrices introduced in the rest of this
section are computed (lines 30--33). In particular, as the algorithm
has to reproduce all the marginal probabilities, we must calculate a
matrix $R$ such that
\begin{equation}
  \nonumber
  {K}\,R\,{K^\prime}={\Lambda} 
\end{equation}
where $K$ is $m \times N$, $K^\prime$ is the transpose conjugate,
$K^+$ is called pseudo-inverse, $R$ is $N \times N$, and $\Lambda$ is
$m \times m$.  For any complex $m \times N$ matrix $K$, the
pseudo-inverse of $K$ is any $N \times m$ matrix $K^+$ such that
\begin{align}
  \label{eq:3}
  KK^+K&=K	& KK^+ &=(KK^+)^\prime \\
  K^+KK^+&=K^+	& K^+K &=(K^+K)^\prime 
\end{align}
where $K^\prime$ is the conjugate transpose of $K$.  One can prove
that $K^+$ is unique and certainly exists.  Moreover, any complex $m
\times N$ matrix of rank $k$ is pseudo-diagonal when it has only zeros
except for $k$ diagonal elements.

For any complex $m \times N$ matrix $K$ of rank $k$, the \ac{SVD} of
$K$ is the product
\begin{equation}
  \label{eq:svd}
  K=V\,S\,U^\prime
\end{equation}
where $V$ is a $m \times m$ unitary matrix, $S$ is a $m \times N$
pseudo-diagonal matrix, and $U$ is a $N \times N$ unitary matrix.  Any
complex $m \times N$ matrix $K$ of rank $k$ admits a \ac{SVD},
although not unique. When a matrix is Hermitean, the \ac{SVD} is an
eigen-decomposition where $V=U$ and $\Sigma$ is a real matrix.

After multiplying both sides of \eqref{eq:svd} by $K^\prime$ on the left and
by $K$ on the right, we obtain
\begin{equation}
  {K}^\prime{K}\,R\,{K^\prime}{K}={K}^\prime\,{\Lambda}\,{K}
  \nonumber
\end{equation}
Let
\begin{equation}
  J={K}^\prime{K}\label{eq:j}
\end{equation}
As $J$ is a Hermitean matrix on a finitely dimensional vector space,
\begin{equation}
  \label{eq:spectral-theorem}
  J=U\,\Sigma\,U^\prime
\end{equation}
where $U=\left(\ket{u}_{1},\dots,\ket{u}_N\right)$ and $\Sigma$ is a
diagonal matrix such that $\mbox{diag}(\Sigma) =
\left(\sigma_{1},\dots,\sigma_N\right)$.  If $\sigma_i \neq \sigma_j$
unless $i=j$ then \eqref{eq:spectral-theorem} is unique.  If
\eqref{eq:spectral-theorem} is applied we have that:
\begin{align*}
  J\,R\,J &={K}^\prime\,{\Lambda}\,{K} \\
  U\,\Sigma\,U^\prime R\,U\,\Sigma\,U^\prime &={K}^\prime {\Lambda}\,{K} \\
  \Sigma\,U^\prime R\,U\,\Sigma &=U^\prime\,{K}^\prime {\Lambda}\,{K}\,U \\
  U^\prime\,R\,U &= \Sigma^+\,U^\prime {K}^\prime {\Lambda}\,{K}\,U\,\Sigma^+       
\end{align*}
and finally
\begin{equation}
  \label{eq:r}
  R=U\,\Sigma^+\,U^\prime\,K^\prime\,{\Lambda}\,K\,U\,\Sigma^+\,U^\prime 
\end{equation}
The marginal probabilities can be restored by computing the following
expression:
\begin{equation}
  \nonumber
  K\,R\,K^\prime 
\end{equation}
Finally, we have that 
\begin{equation}
  \label{eq:rho}
  \rho={\tr({R})^{-1}}\,R 
\end{equation}

The probability function of the \ac{QPS} can also be expressed as a
system of quadratic equations as follows.  Let
\begin{equation}
  W = U\,\Sigma^+\,U^\prime\,K^\prime\nonumber
\end{equation}
and $w_{b,i}$ be an element of $W$. The $b$-th diagonal element of
\eqref{eq:rho} is the probability provided by the \ac{QPS} of the
event represented by $\ket{x}_b$ and can be written as
\begin{equation}
  \label{eq:lin-eq}
  Q(\ket{x}_b)=\sum_{i=1}^N w_{b,i}^2\,\lambda_i
\end{equation}

Note that the number of columns of $K$ grows to an exponential order
of $n$, since there are $N=2^n$ combinations of binary values.
Therefore, the number of variables should then be kept as small as
possible.  In some applications such as \ac{IR} the number of
variables equals the number of query terms, which is usually small.
When $n$ is not small, some heuristics can ameliorate the
computational cost.  Consider the $n(n-1)/2$ marginal bivariate
probabilities and in particular those of events such as $A_1A_2$ and
$A_1\bar{A}_2$.  Each event is represented by a row of $K$ with two
$1$'s and $N-2$ zeros.  As such a row is very sparse, it can be
efficiently stored in compressed format, thus only requiring memory
space for storing the column index of the $1$'s.  Moreover, the first
$n$ rows of $K$ might not be stored since such rows may result from
the join of the corresponding univariate events; for example, the row
of $A_1$ results from the join of the rows of $A_1A_2$ and
$A_1\bar{A}_2$ and can be calculated only if needed.

\subsection{Two Examples of \ac{QPS}}
\label{sec:two-examples}

For starters, consider three binary variables $A_1,A_2,A_3$.  Consider
also their conjunctions $(A_1A_2)$, $(A_1A_3)$, $(A_2A_3)$ in the
sense that each conjunction is measured in one experimental context.
Then, we correspond each triple, such as $A_1A_2A_3$, to a canonical
vector; for example, $\bar{A}_1\bar{A}_2\bar{A}_3$ corresponds to the
vector $\bra{x}_{0}= (1,0,0,0,0,0,0,0)$ and $\bar{A}_1\bar{A}_2A_3$
corresponds to the vector $\bra{x}_{1}=(0,1,0,0,0,0,0,0)$; the join of
the two vectors results in the plane corresponding to
$\bar{A}_1\bar{A}_2$, that is,
\begin{equation}
  \bra{x}_0 \oplus \bra{x}_1 = (1,1,0,0,0,0,0,0)
  \nonumber
\end{equation}
and ${\bra{x}_{0} \oplus \bra{x}_{1} \oplus \bra{x}_{2} \oplus
  \bra{x}_{3}}$ will be vector
\begin{equation}
  \nonumber
  \begin{pmatrix}
    1 & 1 & 1 & 1 & 0 & 0 & 0 & 0
  \end{pmatrix}
\end{equation}
If the join is repeated for all the events corresponding to the six
measured variables, the following matrix is obtained
\begin{equation}
  \nonumber
  {K}=
  \begin{pmatrix}
    1 & 1 & 1 & 1 & 0 & 0 & 0 & 0\\
    1 & 1 & 0 & 0 & 1 & 1 & 0 & 0\\
    1 & 0 & 1 & 0 & 1 & 0 & 1 & 0\\
    0 & 0 & 0 & 0 & 0 & 0 & 1 & 1\\
    0 & 0 & 0 & 0 & 0 & 1 & 0 & 1\\
    0 & 0 & 0 & 1 & 0 & 0 & 0 & 1
  \end{pmatrix}
\end{equation}
where $m$ is the number of marginal probabilities and $N$ is the
number of basis vectors.  The first row corresponds to $P(\bar{A}_1)$
because it is obtained by the join of the subspaces representing
$\bar{A}_1\bar{A}_2\bar{A}_3$, $\bar{A}_1\bar{A}_2A_3$,
$\bar{A}_1A_2\bar{A}_3$ and $\bar{A}_1A_2A_3$. It can be easily
checked that, if $K[i]$ is the $i$-th row of $K$ we have that
\begin{align*}
  K[1] \,\rho\,K[1]^\prime&=P(\bar{A}_1)&K[4] \,\rho\,K[4]^\prime&=P(A_1A_2)\\
  K[2] \,\rho\,K[2]^\prime&=P(\bar{A}_2)&K[5] \,\rho\,K[5]^\prime&=P(A_1A_3)\\
  K[3] \,\rho\,K[3]^\prime&=P(\bar{A}_3)&K[6] \,\rho\,K[6]^\prime&=P(A_2A_3)
\end{align*}
Therefore, the problem of calculating one probability space can be
stated as the problem of calculating a density matrix $\rho$ such that
\begin{align*}
  P(\bar{A}_1)	 &=\left(\bra{x}_{0} \oplus \bra{x}_{1} \oplus \bra{x}_{2} \oplus \bra{x}_{3}\right) \,\rho\, \left(\ket{x}_{0} \oplus \ket{x}_{1} \oplus \ket{x}_{2} \oplus \ket{x}_{3}\right)\\
  P(\bar{A}_2)	 &=\left(\bra{x}_{0} \oplus \bra{x}_{1} \oplus \bra{x}_{4} \oplus \bra{x}_{5}\right) \,\rho\, \left(\ket{x}_{0} \oplus \ket{x}_{1} \oplus \ket{x}_{4} \oplus \ket{x}_{5}\right)\\
  P(\bar{A}_3) 	 &=\left(\bra{x}_{0} \oplus \bra{x}_{2} \oplus \bra{x}_{4} \oplus \bra{x}_{6}\right) \,\rho\, \left(\ket{x}_{0} \oplus \ket{x}_{2} \oplus \ket{x}_{4} \oplus \ket{x}_{6}\right) \\
  P(A_1A_2) &=\left(\bra{x}_{6} \oplus \bra{x}_{7}\right) \,\rho\, \left(\ket{x}_{6} \oplus \ket{x}_{7}\right) \\
  P(A_1A_3) &=\left(\bra{x}_{5} \oplus \bra{x}_{7}\right) \,\rho\, \left(\ket{x}_{5} \oplus \ket{x}_{7}\right)\\
  P(A_2A_3) &=\left(\bra{x}_{3} \oplus \bra{x}_{7}\right) \,\rho\, \left(\ket{x}_{3} \oplus \ket{x}_{7}\right)
\end{align*}
and
\begin{equation}
  \nonumber
  0 \leq \bra{x}_b\,\rho\,\ket{x}_b \leq 1 \quad\mbox{for all}\quad b
  \in \{0,1\}^3 \quad\mbox{such that}\quad \sum_{b \in \{0,1\}^3}
  \bra{x}_b\,\rho\,\ket{x}_b=1 
\end{equation}
where $\rho = R/\tr{(R)}$.  A numerical example is provided in the
following.  Suppose
\begin{equation}
  \nonumber
  \Lambda=\mbox{diag}\left( \frac{1}{2}, \frac{1}{2}, \frac{1}{2},
    \frac{9}{20}, \frac{9}{20}, \frac{1}{10} \right)
\end{equation}
which violates \eqref{eq:inequality} and then does not admit a
set-based probability space.  We have that the probability distributed
along the diagonal of $\rho$ is
\begin{equation}
  \nonumber
  \left\{ 0.0105,0.242,0.242,0.0469,0.201,0.115,0.115,0.0274 \right \}
\end{equation}
The sum of the first four values is $P(\bar{A}_1)$.  Moreover, suppose
\begin{equation}
  \nonumber
  \Lambda=\mbox{diag}\left( \frac{1}{2}, \frac{1}{2}, \frac{1}{2},
    \frac{1}{20}, \frac{9}{20}, \frac{1}{10} \right)
\end{equation} 
which does not violate \eqref{eq:inequality}.  Accordingly, $P(\ket{x}_1\ket{x}_2\ket{x}_3)= \frac{1}{20}$
in a set-based probability space.  We have that the probabilities
distributed along the diagonal of $\rho$ are
\begin{equation}
  \nonumber
  \left\{ 0.0134, 0.227, 0.287, 0.0469, 0.234, 0.135, 0.0343, 0.0217 \right \}
\end{equation}
Therefore, the set-based distribution is not necessarily reproduced
although \eqref{eq:inequality} holds.  However, the marginal
probabilities have again been restored even in the new space:
\begin{align*}
  K[1] \,\rho\,K[1]^\prime  &=0.50 &K[4] \,\rho\,K[4]^\prime &=0.05\\
  K[2] \,\rho\,K[2]^\prime  &=0.50 &K[5] \,\rho\,K[5]^\prime &=0.45\\
  K[3] \,\rho\,K[3]^\prime  &=0.50 &K[6] \,\rho\,K[6]^\prime &=0.10
\end{align*}
Using \eqref{eq:lin-eq}, an alternative expression of the probability
space can be provided in terms of equations as follows:
\begin{align*}
  P(\bar{A}_1\bar{A}_2\bar{A}_3)	&= \frac{0.02 \lambda_1+0.02 \lambda_2+0.02 \lambda_3+0.001 \lambda_4+0.001 \lambda_5+0.001 \lambda_6}{0.7 \lambda_1+0.7 \lambda_2+0.7 \lambda_3+ \lambda_4+ \lambda_5+ \lambda_6}\\
  P(\bar{A}_1\bar{A}_2A_3) 		&= \frac{0.2 \lambda_1+0.2 \lambda_2+0.3 \lambda_3+0.4 \lambda_4+0.1 \lambda_5+0.1 \lambda_6}{0.7 \lambda_1+0.7 \lambda_2+0.7 \lambda_3+ \lambda_4+ \lambda_5+ \lambda_6}         \\
  P(\bar{A}_1A_2\bar{A}_3) 		&= \frac{0.2 \lambda_1+0.3 \lambda_2+0.2 \lambda_3+0.1 \lambda_4+0.4 \lambda_5+0.1 \lambda_6}{0.7 \lambda_1+0.7 \lambda_2+0.7 \lambda_3+ \lambda_4+ \lambda_5+ \lambda_6}         \\
  P(\bar{A}_1A_2A_3) 				&= \frac{0.001 \lambda_1+0.001 \lambda_2+0.001 \lambda_3+0.07 \lambda_4+0.07 \lambda_5+0.6 \lambda_6}{0.7 \lambda_1+0.7 \lambda_2+0.7 \lambda_3+ \lambda_4+ \lambda_5+ \lambda_6} \\
  P(A_1\bar{A}_2\bar{A}_3) 		&= \frac{0.3 \lambda_1+0.2 \lambda_2+0.2 \lambda_3+0.1 \lambda_4+0.1 \lambda_5+0.4 \lambda_6}{0.7 \lambda_1+0.7 \lambda_2+0.7 \lambda_3+ \lambda_4+ \lambda_5+ \lambda_6}         \\
  P(A_1\bar{A}_2A_3) 				&= \frac{0.001 \lambda_1+0.001 \lambda_2+0.001 \lambda_3+0.07 \lambda_4+0.6 \lambda_5+0.07 \lambda_6}{0.7 \lambda_1+0.7 \lambda_2+0.7 \lambda_3+ \lambda_4+ \lambda_5+ \lambda_6} \\
  P(A_1A_2\bar{A}_3) 				&= \frac{0.001 \lambda_1+0.001 \lambda_2+0.001 \lambda_3+0.6 \lambda_4+0.07 \lambda_5+0.07 \lambda_6}{0.7 \lambda_1+0.7 \lambda_2+0.7 \lambda_3+ \lambda_4+ \lambda_5+ \lambda_6} \\
  P(A_1A_2A_3) 					&= \frac{0.001 \lambda_1+0.001 \lambda_2+0.001 \lambda_3+0.07 \lambda_4+0.07 \lambda_5+0.07 \lambda_6}{0.7 \lambda_1+0.7 \lambda_2+0.7 \lambda_3+ \lambda_4+ \lambda_5+ \lambda_6}
\end{align*}
In the event of 4 binary observables, we have the following: 
\begin{equation*}
  K=\left(
    \begin{array}{cccccccccccccccc}
      1&1&1&1&1&1&1&1&0&0&0&0&0&0&0&0\\
      1&1&1&1&0&0&0&0&1&1&1&1&0&0&0&0\\
      1&1&0&0&1&1&0&0&1&1&0&0&1&1&0&0\\
      1&0&1&0&1&0&1&0&1&0&1&0&1&0&1&0\\
      0&0&0&0&0&0&0&0&0&0&0&0&1&1&1&1\\
      0&0&0&0&0&0&0&0&0&0&1&1&0&0&1&1\\
      0&0&0&0&0&0&0&0&0&1&0&1&0&1&0&1\\
      0&0&0&0&0&0&1&1&0&0&0&0&0&0&1&1\\
      0&0&0&0&0&1&0&1&0&0&0&0&0&1&0&1\\
      0&0&0&1&0&0&0&1&0&0&0&1&0&0&0&1
    \end{array}
  \right)
\end{equation*}
and
\begin{footnotesize}
  \begin{equation*}
    \mbox{diag}(\Lambda) = \left(P(\bar{A}_1), P(\bar{A}_2),
      P(\bar{A}_3), P(\bar{A}_4), P(A_1A_2), P(A_1A_3), P(A_1A_4),
      P(A_2A_3),  P(A_2A_4), P(A_3A_4)\right)
  \end{equation*}
\end{footnotesize}

\noindent 
The diagonal of $R$ is the distribution of probability of
$A_1,A_2,A_3,A_4$.  The linear equation system can be written as
follows:
\begin{footnotesize}
  \begin{align*}
    P(\bar{A}_1\bar{A}_2\bar{A}_3\bar{A}_4)	&=\frac{0.003 (\lambda_1+ \lambda_2+ \lambda_3+ \lambda_4+ \lambda_5+ \lambda_6+ \lambda_7+ \lambda_8+ \lambda_9+ \lambda_{10})}{0.6 (\lambda_1+ \lambda_2+ \lambda_3+ \lambda_4)+0.8 (\lambda_5+ \lambda_6+ \lambda_7+ \lambda_8+ \lambda_9+ \lambda_{10})}\\
    P(\bar{A}_1\bar{A}_2\bar{A}_3A_4)			&=\frac{0.05 (\lambda_1+ \lambda_2+ \lambda_3)+0.3 \lambda_4+0.06 (\lambda_5+ \lambda_6)+0.07 \lambda_7+0.06 \lambda_8+0.07 (\lambda_9+ \lambda_{10})}{0.6 (\lambda_1+ \lambda_2+ \lambda_3+ \lambda_4)+0.8 (\lambda_5+ \lambda_6+ \lambda_7+ \lambda_8+ \lambda_9+ \lambda_{10})}\\
    P(\bar{A}_1\bar{A}_2A_3\bar{A}_4)			&=\frac{0.05 (\lambda_1+ \lambda_2)+0.3 \lambda_3+0.05 \lambda_4+0.06 \lambda_5+0.07 \lambda_6+0.06 \lambda_7+0.07 \lambda_8+0.06 \lambda_9+0.07 \lambda_{10}}{0.6 (\lambda_1+ \lambda_2+ \lambda_3+ \lambda_4)+0.8 (\lambda_5+ \lambda_6+ \lambda_7+ \lambda_8+ \lambda_9+ \lambda_{10})}\\
    P(\bar{A}_1\bar{A}_2A_3A_4)				&=\frac{0.02 (\lambda_1+ \lambda_2)+0.01 (\lambda_3+ \lambda_4)+0.1 \lambda_5+0.03 (\lambda_6+ \lambda_7+ \lambda_8+ \lambda_9)+0.1 \lambda_{10}}{0.6 (\lambda_1+ \lambda_2+ \lambda_3+ \lambda_4)+0.8 (\lambda_5+ \lambda_6+ \lambda_7+ \lambda_8+ \lambda_9+ \lambda_{10})}\\
    P(\bar{A}_1A_2\bar{A}_3\bar{A}_4)			&=\frac{0.05 \lambda_1+0.3 \lambda_2+0.05 (\lambda_3+ \lambda_4)+0.07 \lambda_5+0.06 (\lambda_6+ \lambda_7)+0.07 (\lambda_8+ \lambda_9)+0.06 \lambda_{10}}{0.6 (\lambda_1+ \lambda_2+ \lambda_3+ \lambda_4)+0.8 (\lambda_5+ \lambda_6+ \lambda_7+ \lambda_8+ \lambda_9+ \lambda_{10})}\\
    P(\bar{A}_1A_2\bar{A}_3A_4)				&=\frac{0.02 \lambda_1+0.01 \lambda_2+0.02 \lambda_3+0.01 \lambda_4+0.03 \lambda_5+0.1 \lambda_6+0.03 (\lambda_7+ \lambda_8)+0.1 \lambda_9+0.03 \lambda_{10}}{0.6 (\lambda_1+ \lambda_2+ \lambda_3+ \lambda_4)+0.8 (\lambda_5+ \lambda_6+ \lambda_7+ \lambda_8+ \lambda_9+ \lambda_{10})}\\
    P(\bar{A}_1A_2A_3\bar{A}_4)				&=\frac{0.02 \lambda_1+0.01 (\lambda_2+ \lambda_3)+0.02 \lambda_4+0.03 (\lambda_5+ \lambda_6)+0.1 (\lambda_7+ \lambda_8)+0.03 (\lambda_9+ \lambda_{10})}{0.6 (\lambda_1+ \lambda_2+ \lambda_3+ \lambda_4)+0.8 (\lambda_5+ \lambda_6+ \lambda_7+ \lambda_8+ \lambda_9+ \lambda_{10})}\\
    P(\bar{A}_1A_2A_3A_4)						&=\frac{0.03 \lambda_1+0.004 (\lambda_2+ \lambda_3+ \lambda_4)+0.04 (\lambda_5+ \lambda_6+ \lambda_7)+0.08 (\lambda_8+ \lambda_9+ \lambda_{10})}{0.6 (\lambda_1+ \lambda_2+ \lambda_3+ \lambda_4)+0.8 (\lambda_5+ \lambda_6+ \lambda_7+ \lambda_8+ \lambda_9+ \lambda_{10})}\\
    P(A_1\bar{A}_2\bar{A}_3\bar{A}_4)			&=\frac{0.3 \lambda_1+0.05 (\lambda_2+ \lambda_3+ \lambda_4)+0.07 (\lambda_5+ \lambda_6+ \lambda_7)+0.06 (\lambda_8+ \lambda_9+ \lambda_{10})}{0.6 (\lambda_1+ \lambda_2+ \lambda_3+ \lambda_4)+0.8 (\lambda_5+ \lambda_6+ \lambda_7+ \lambda_8+ \lambda_9+ \lambda_{10})}\\
    P(A_1\bar{A}_2\bar{A}_3A_4)				&=\frac{0.01 \lambda_1+0.02 (\lambda_2+ \lambda_3)+0.01 \lambda_4+0.03 (\lambda_5+ \lambda_6)+0.1 (\lambda_7+ \lambda_8)+0.03 (\lambda_9+ \lambda_{10})}{0.6 (\lambda_1+ \lambda_2+ \lambda_3+ \lambda_4)+0.8 (\lambda_5+ \lambda_6+ \lambda_7+ \lambda_8+ \lambda_9+ \lambda_{10})}\\
    P(A_1\bar{A}_2A_3\bar{A}_4)				&=\frac{0.01 \lambda_1+0.02 \lambda_2+0.01 \lambda_3+0.02 \lambda_4+0.03 \lambda_5+0.1 \lambda_6+0.03 (\lambda_7+ \lambda_8)+0.1 \lambda_9+0.03 \lambda_{10}}{0.6 (\lambda_1+ \lambda_2+ \lambda_3+ \lambda_4)+0.8 (\lambda_5+ \lambda_6+ \lambda_7+ \lambda_8+ \lambda_9+ \lambda_{10})}\\
    P(A_1\bar{A}_2A_3A_4)						&=\frac{0.004 \lambda_1+0.03 \lambda_2+0.004 (\lambda_3+ \lambda_4)+0.04 \lambda_5+0.08 (\lambda_6+ \lambda_7)+0.04 (\lambda_8+ \lambda_9)+0.08 \lambda_{10}}{0.6 (\lambda_1+ \lambda_2+ \lambda_3+ \lambda_4)+0.8 (\lambda_5+ \lambda_6+ \lambda_7+ \lambda_8+ \lambda_9+ \lambda_{10})}\\
    P(A_1A_2\bar{A}_3\bar{A}_4)				&=\frac{0.01 (\lambda_1+ \lambda_2)+0.02 (\lambda_3+ \lambda_4)+0.1 \lambda_5+0.03 (\lambda_6+ \lambda_7+ \lambda_8+ \lambda_9)+0.1 \lambda_{10}}{0.6 (\lambda_1+ \lambda_2+ \lambda_3+ \lambda_4)+0.8 (\lambda_5+ \lambda_6+ \lambda_7+ \lambda_8+ \lambda_9+ \lambda_{10})}\\
    P(A_1A_2\bar{A}_3A_4)						&=\frac{0.004 (\lambda_1+ \lambda_2)+0.03 \lambda_3+0.004 \lambda_4+0.08 \lambda_5+0.04 \lambda_6+0.08 \lambda_7+0.04 \lambda_8+0.08 \lambda_9+0.04 \lambda_{10}}{0.6 (\lambda_1+ \lambda_2+ \lambda_3+ \lambda_4)+0.8 (\lambda_5+ \lambda_6+ \lambda_7+ \lambda_8+ \lambda_9+ \lambda_{10})}\\
    P(A_1A_2A_3\bar{A}_4)						&=\frac{0.004 (\lambda_1+ \lambda_2+ \lambda_3)+0.03 \lambda_4+0.08 (\lambda_5+ \lambda_6)+0.04 \lambda_7+0.08 \lambda_8+0.04 (\lambda_9+ \lambda_{10})}{0.6 (\lambda_1+ \lambda_2+ \lambda_3+ \lambda_4)+0.8 (\lambda_5+ \lambda_6+ \lambda_7+ \lambda_8+ \lambda_9+ \lambda_{10})}\\
    P(A_1A_2A_3A_4) &=\frac{\left(0.0006 (\lambda_1+ \lambda_2+
        \lambda_3+ \lambda_4)+0.007 (\lambda_5+ \lambda_6+ \lambda_7+
        \lambda_8+ \lambda_9+ \lambda_{10})\right)}{0.6 (\lambda_1+
      \lambda_2+ \lambda_3+ \lambda_4)+0.8 (\lambda_5+ \lambda_6+
      \lambda_7+ \lambda_8+ \lambda_9+ \lambda_{10})}
  \end{align*}
\end{footnotesize}
An implementation of the \ac{QPS}'s algorithm can be utilized at
\begin{center}
  \url{http://isotta.dei.unipd.it/cgi-bin/qps/qps-w-form.py}
\end{center}
by using any $3\leq n\leq 14$.

\end{document}